\documentstyle[preprint,revtex,eqsecnum]{aps}

\begin{document}
\begin{flushright}
BA-93-07
\end{flushright}
\vskip0.5truein
\draft
\begin{title}
Calculated Electron Fluxes at Airplane Altitudes
\end{title}
\author{ R.~K.~Schaefer, T.~K.~Gaisser, and T.~S.~Stanev}
\begin{instit}
 Bartol Research Institute \\
 University of Delaware, Newark, DE 19716\\
\end{instit}
\receipt{22 Feb 1993}

\begin{abstract}

A precision measurement of atmospheric electron fluxes has been performed
on a Japanese commercial airliner (Enomoto, {\it et al.}, - Ref.
\cite{enomoto}).  The bulk of these electrons are produced in pairs from the
gamma rays emitted when $\pi^0$s decay, which in turn have been produced in
cosmic ray-air nucleus collisions.  These electron fluxes can be used to test
elements of our atmospheric neutrino flux calculation, {\it i.e.}, the assumed
primary spectrum and the monte carlo shower code.
Here we have modified the monte carlo program which
had previously been used to calculate the fluxes of atmospheric neutrinos by
combining it with the program GEANT to compute the electromagnetic part of the
shower.  This hybrid program now keeps track of the electrons produced in
cosmic ray showers as a function of energy and atmospheric depth.  We compare
our calculated integral fluxes above the experimental threshold energies
$1$ GeV, $2$ GeV, and $4$ GeV for a variety of atmospheric depths and cutoff
rigidities.  Our results are in good agreement ($\sim$ a few \%) with the
data, but we found we needed to boost the normalization of the primary flux
by 12\% over the value we had previously used to calculate the atmospheric
neutrino flux.

\end{abstract}


\narrowtext

\section{Introduction}
\label{sec:level1}

The production of secondary electrons in the atmosphere from primary
cosmic rays is a well known phenomenon and a detailed formalism for
calculating these effects has existed for a long time.  Most
measurements of electron fluxes have been carried out with balloons which
spend most of their time near the top of the atmosphere.  Thus, most
balloon measurements reflect cosmic rays which have penetrated $\sim 2-50$
gm/cm$^2$ of atmosphere.  Larger depths are not well sampled by balloons.
To reach the altitudes of commercial jet aircraft, $\sim 9-12$ km, a cosmic
ray shower has to penetrate
${\mathrel{\raise.3ex\hbox{$>$\kern-.75em\lower1ex\hbox{$\sim$}}}} 200$
gm/cm$^2$ of (vertical) material.
Since this corresponds to  about 6  electron radiation lengths, there are
virtually no primary cosmic ray  electrons which can penetrate to this
depth.  The bulk of the atmospheric electrons detected in an airplane would
then be secondaries generated in  showers from  primary cosmic ray
nucleons.  As such, these electrons offer us a probe of the primary cosmic
ray spectrum, and high precision measurements could potentially pin down the
proper normalization of the high energy cosmic ray flux, which is
uncertain (see {\it e.g.,\ } Ref. \cite{gs}) by about $\pm 20$ \%.

    In fact this uncertainty in the normalization of the high energy cosmic
ray spectrum affects all calculations of cosmic ray progeny, including gamma
rays, antiprotons, positrons, and atmospheric neutrinos.  For example, in the
recent measurements of the atmospheric neutrinos, one does not know whether
the detectors are seeing too few muon neutrinos or too many electron
neutrinos because the expected rate depends directly on how the calculation is
normalized \cite{gaisser}.

    Early treatments of electron production (see, Ref. \cite{DanielS})
consisted of integrating a set of coupled atmospheric cascade
equations.   However, recent calculations of the atmospheric neutrino fluxes
have been done using a detailed
monte carlo program for cosmic ray air showers developed over a period of
time.  It is therefore useful, convenient, and relevant that we use this same
program to calculate the electron fluxes expected in an airplane
experiment.  We have two main objectives in doing the calculation this way:
1) We hope to test the accuracy of the monte carlo program, and 2) we hope
to fix the normalization of the primary nucleon energy spectrum.  The
results are also relevant to calibration of balloon-borne electron
spectrometers.  These must measure the electron flux as the balloon rises
in order to subtract the secondary component at float altitude.  A precise
calculation of flux vs. altitude is helpful in making this subtraction.
In the next section we describe how we performed the
calculation, and then we will discuss our results and implications.

\section{Calculations}

  The primary goal of our computation is to compare the fluxes
measured in the Enomoto, {\it et al.}, (Ref. \cite{enomoto}) experiment.
This procedure is quite similar to one we did for the atmospheric
neutrino flux \cite{Barrgs,BGS}. We will therefore only
briefly summarize the relevant parts of the calculation.

   We begin by simulating cascades initiated by a primary particle of fixed
energy.  (An incoming nucleus of mass $A$ is treated as $A$ independent
incident nucleons - an approximation which has been shown to correctly
reproduce the uncorrelated atmospheric average fluxes of particles
\cite{Engel}.)  The primary nucleon collides with an air nucleus
producing a fast fragment nucleon and many  mesons, of which some decay and
some interact again.  All $\pi^0$ mesons produced along the shower core decay
essentially immediately into a pair of photons which then
subsequently produce negatron-positron pairs.  Occasionally other mesons
will decay directly to an electron (e.g.,
$K^0_L \rightarrow \pi^\pm e^\mp \nu$ or $K^+ \rightarrow \pi^0 e^+ \nu_e$).
As these electrons propagate in the atmosphere, they can emit bremstrahlung
photons which later produce more negatron-positron pairs.  Bremstrahlung is
the main source of energy loss for electrons in this energy range, and the
calculated flux of electrons at a given depth is quite sensitive to
the accuracy of the description of the bremstrahlung process.  In order
to treat all electromagnetic effects (including bremstrahlung) to a high degree
of accuracy we have used the program GEANT, a standard code used by high energy
particle physics experimenters for calculating the particle showers that
happen inside their detectors.

     GEANT incorporates many electromagnetic effects which are not very
important at GeV energies, but can affect the results at the few percent
level.    We are striving for high precision, so we used GEANT to calculate the
following effects as well as the bremstrahlung: ionization energy losses in
air, Compton scattering of atomic electrons, delta ray production, and multiple
scattering.  GEANT does its simulations using constant density slabs of
material, unlike our hadronic code which uses a continuously varying air
density.  The electromagnetic processes are not sensitive to the depth/altitude
relationship, but the hadronic processes are, (especially to the competition
between decay and interaction for charged pions and kaons).  Thus we achieved
the following
synthesis: first, we simulate the hadronic portion of the shower in a realistic
atmospheric model, keeping track of the
depths and energies of all of the produced photons (and electrons from direct
meson decays).  At each primary energy, we simulate as many
showers as it takes to accumulate $>50000$ particles (mostly photons with a few
direct $e^\pm$) with energies greater than 1 GeV at altitudes above the minimum
for which the Enomoto {\it et al.}, \cite{enomoto} data was taken.  Then
we input these photons (and electrons) into the GEANT program which uses a
constant density  ($\rho_{air} = 1.19\times 10^{-4}$ gm/cm$^3$) slab of air.
We then record the number of electrons which pass 20 evenly
spaced depth flags between 205 and 300 gm/cm$^2$.  We repeat this process
for 22 separate (roughly logarithmically spaced) primary energies for each of
the three electron threshold energies 1, 2, and 4 GeV.  We estimate the
statistical uncertainty in our Monte Carlo contributes $\leq1$ \% uncertainty
to the final electron fluxes.

  The yields per incident primary are then multiplied by the differential
primary proton spectrum to get the electron yield per primary energy
interval.  The primary proton spectrum used here has the same shape as that
used in Ref. \cite{BGS}, but with a 12\% higher normalization.
Multiplying this  quantity by the
energy gives the electron yield per {\sl logarithmic} energy interval which is
plotted in figure 1 at a typical  airplane depth (250 gm/cm$^2$) for each
of the electron energy thresholds 1, 2, and 4 GeV.  One can see that the
electron flux drops sharply with energy and that the
average primary energy which contributes these
electrons rises proportionately with the threshold energy.   The primaries
responsible for the electrons in these measurements are of somewhat higher
energy than those responsible for the atmospheric neutrinos measured at IMB
\cite{casper} and Kamiokande \cite{hirata}.  For example,
the median primary energy for
progenitors of GeV neutrinos is approximately 20 GeV/nucleon, as compared
to 50 GeV/nucleon for GeV electrons.  The difference is a consequence of
the electromagnetic cascading process.

To first approximation, the total electron number which would be measured at
250 gm/cm$^2$ is just the area under the curve in figure 1.  However, in
contrast to the typical balloon experiment, the airplane experiment measured
fluxes in locations where the geomagnetic primary cutoff energies are quite
high (rigidities in the range 4.8-16.2 GV) and at a variety of
atmospheric depths.  These effects must be treated accurately.  The cutoff
rigidities were handled as follows.  For protons, the cutoff rigidities are
converted to energies, and the integration of the primary spectrum is
cutoff abruptly at this energy.  For primary nuclei, the cutoff energy per
nucleon will be different (roughly a factor of 2 lower than for protons) so
we include the primary nuclei separately.

The detector has an acceptance of 15$^\circ$ from vertical.  To simulate
particles which are incident to the atmosphere at an angle $\theta_{incident}$
from vertical, one must scale all of the depths by a factor
$1/cos(\theta_{incident})$.  We have explicitly done an integration over the
solid angle contained in a $15^\circ$ cone in a few test cases and find the
results are indistinguishable from simply scaling all of the depths to the
average angle, $10.6^\circ$, and performing no additional averaging.  Since
$1/cos(10.6^\circ)= 1.02$, the slant depth is 2\% larger than vertical.

Finally, we have considered secondary electrons generated from primary
$electrons$.  These were not included in the treatment of Ref. \cite{DanielS}.
 The primary electrons have lost enough energy via
bremstrahlung that they do not themselves contribute to the flux at airplane
altitudes.  However, the bremstrahlung photons these primary electrons emit
can also produce negatron-positron pairs after penetrating a bit
deeper than the primary electrons.
We have used the Monte carlo program GEANT with primary electrons to simulate
the electromagnetic cascades generated by primary electrons.   For purposes
of including electronic
progenitor electrons in our airplane flux calculation, we assume that the
primary electrons, although of opposite charge sign, will
have the same cutoff rigidity as the protons.  This approximation is
adequate because cascading insures that most of the contribution to electrons
$>1$ GeV at $\geq200$ gm/cm$^2$ comes from primary electrons above the
geomagnetic cutoff.  For the primary electron
spectrum we use the spectrum from the 1990 flight of the LEE (Low Energy
Electron) balloon experiment
(Ref. \cite{Evenson}), although using a different spectrum (1987 LEE flight)
gave results indistinguishable from those presented here.  When we include
these secondary
calculations into our results the fluxes increase by about 3\% for $>1$ GeV
electrons and somewhat less for higher threshold energies.

The total calculation can be represented by the following formula:
\begin{equation}
 N_e(>E_{thr}, d_{atm}, R_c) = \sum_{i=1}^3\int_{E_i(R_c)}^\infty dE
{dN_i\over dE} Y_i(E, E_{thr}, 1.02 d_{atm})
\end{equation}
Here $N_e$ is the predicted integral electron flux above threshold energy
$E_{thr}$, vertical atmospheric depth $d_{atm}$, and cutoff rigidity $R_c$.  We
have
to sum over the different types of primaries: protons, electrons, and
nuclei (which have a different cutoff energy than protons, hence the lower
limit on the integral $E_i(R_c)$ depends on primary type). The $Y_i$ are
the electron yields  per incident primary of energy E and type $i$ at a
depth $d_{atm}$ and electron threshold  energy $E_{thr}$.  The 1.02 factor
multiplying $d_{atm}$ in the yield is to account for the fact that the
average primary is incident at an angle of $10.6^\circ$ as described above.

   The results of these calculations are shown by the triangles in
figures 2, 3, and 4 for energy thresholds of 1, 2, and 4 GeV respectively.
The measurements and error bars of Enomoto, {\it et al.}, \cite{enomoto} are
presented for comparison.  These figures correspond directly to Figure 5 of
Enomoto, {\it et al.}, but we have placed the data from both the Guam-Sydney
and the Sidney-Guam flights on the same graph. Thus the data seemingly taken at
nearby latitudes can in fact have been taken at quite different altitudes and
cutoff rigidities.  We also note here that the
Enomoto {\it et al.} error bars are a quadrature sum of 3\% statistical and
5\% systematic errors.

\section{Discussion and Implications}

    Inspection of the figures 2, 3, and 4 reveals that our calculation
agrees well with the experimental results.  In the plots the triangles
represent our calculations and the filled boxes (with 1 $\sigma$ error bars)
are the measurements of Enomoto, {\it et al.} \cite{enomoto}.  The points are
taken at a variety of atmospheric depths and cutoff rigidities as the
aircraft flew from high to low (terrestrial) latitude and back.  We have
chosen to plot the data as a
function of latitude because this keeps the data points clearly separated.
However, this choice of plotting does not adequately reflect the fact that
points similar latitudes can correspond to data taken at
quite different altitudes and cutoff rigidities, which have been incorporated
explicitly into the calculated fluxes.  (For the altitudes and cutoff
rigidities associated with observations at each latitude point, see Tables
III, IV, and V in Enomoto, {\it et al.}, \cite{enomoto}).
The data tend to be clumped around several depths and cutoff rigidities.
The value of the flux is most sensitive to atmospheric depth and the highest
intensities are recorded at depths of about 220 gm/cm$^2$.  The dependence
of intensity on cutoff rigidity is less sensitive, although still noticeable.
We can see that the general trends with
depth and geomagnetic cutoff are quite well represented to get such good
agreement.

  From Figure 2, we can see the $>1$ GeV electron results are in
remarkably good agreement with the predictions.   Only 2 out of the
18 predictions are not within the
$\pm 1 \sigma$ error bars of the measurements, indicating perhaps that the
experimental errors are slightly overestimated.  Similarly the agreement in
figures 3 and 4 is also better than would be expected from the size of the
errors.   For the $>2$ GeV electron measurements, 4 out of the 15 points are
not within $\pm 1 \sigma$, while for the $>4$ GeV there are only 2 out of 15
outlying points.   A change of the primary flux normalization by just a
few percent ruins this close agreement.

   This calculation shows that our monte carlo program for simulating hadronic
cosmic ray showers seems to working quite well.  The only feature of our
calculation which we could consider improving is the slight systematic
underprediction of the highest altitude flux.  (These points
are found at the upper left corner of all of the figures.)   Coincidently,
these highest altitude points are also those at the lowest rigidity cutoff.  If
this happened only with the $>1$ GeV electrons one might believe this could be
caused by solar modulation of the lower energy flux.  However this
underprediction shows up in the $>2$ and $>4$ GeV electrons as well.  A check
of the differential fluxes (Figure 1) shows that almost none of the
$>4$ GeV electrons come from primaries which are even close to the geomagnetic
cutoff energies.  Thus, there may be a slight problem in that the showers do
not develop high enough in the atmosphere.   This discrepancy is at a level of
5\%.

  We have not included the contribution from ``knock-on" electrons.  The
main source of knock-on electrons at depths $>100$ gm/cm$^2$ is collisions
of muons with air atoms (see Ref. \cite{DanielS}).
Using the formula in Daniel and Stephens (\cite{DanielS}) for producing
knock-ons and the measured muon flux at 9 km
\cite{Baradzei} we estimate that knock-ons could contribute no more than
1\% of the electron flux.

    The excellent fit to the data requires that the normalization of the
primary flux be fixed to within a few percent.  Similarly, changing the
spectral exponent of the primary flux will destroy the agreement of the $>4$
GeV electron data relative to the $>1$ GeV data.  Thus the primary spectrum
used in \cite{BGS} for the calculation of the neutrino
flux seems to have the correct shape, but the normalization should be raised
by 12\%.  If this raised normalization extends to the lower energy part of the
primary spectrum, this would imply that the corresponding neutrino fluxes
should also be scaled up by 12\%.  We note however, that there are systematic
experimental errors which could affect this conclusion.  Energy dependent
uncertainties exist in modeling the acceptance and particle energy
identification \cite{Enoemail}.   Since these errors are
of order 5\%, the integral fluxes could change at worst by as much as
$(1.05)^{1.7} = 1.09$ or 9 \%.  To be very conservative we could therefore
say that the electron measurements imply a normalization
boost of 12 $\pm 9$ \%.  We are currently investigating other indirect
checks (e.g. atmospheric $\mu$ fluxes) to test the consistency of raising the
normalization.  We note that these electron fluxes imply tighter constraints
on the primary spectrum than are currently allowed by direct measurements of
the primary spectrum ($\sim \pm 20$ \%).

\section{ Conclusions}

    We have performed a monte carlo calculation of the atmospheric electron
flux at the depths and cutoff rigidities of the Enomoto, {\it et al.},
airplane experiment.  We predict integral fluxes which show
excellent agreement with the $>1$, $>2$, and $>4$ GeV electrons.  The
predictions generally follow the depth and cutoff rigidity trends seen in the
data.  The agreement is so good that it suggests that the experimental errors
have been estimated very conservatively.
This analysis suggests three things: 1) the hadronic monte carlo code used for
calculating atmospheric neutrino fluxes seems to work well for another species
of atmospheric secondaries, namely electrons, and 2) the shape of the
primary spectrum from Ref. \cite{BGS} gives the correct energy spectrum of
electrons from 1 GeV to 4 GeV, and 3) the normalization of the primary cosmic
ray spectrum is somewhat higher than that used by Ref. \cite{BGS}.

\acknowledgments

We thank R. Enomoto, P. Evenson, M. Honda, and T. Kifune for helpful
discussions.  We would also like to thank J. Petrakis for invaluable help in
using GEANT.  This work is supported in part by NASA grant NAGW-1644.

\newpage

\figure{ Differential electron flux.  The number of electrons (cm$^2$
s sr)$^{-1}$ per logarithmic energy interval as a function of total energy per
nucleon. note the increase in median primary energy and decrease in flux as
the we increase the detector threshold energy.\label{deflx}}

\figure{ Integral Flux at $>$ 1 GeV.  The electron flux (cm$^2$ s
sr)$^{-1}$ predictions for the cutoff rigidities and atmospheric depths
given in Enomoto, {\it et al.}, (1991) - Ref. \cite{enomoto}.  We have plotted
both the Guam-Sydney and the return flight together here as a function of the
terrestrial latitude of the observation.  The apparent jitter in the fluxes is
due to the fact that observations at neighboring latitudes can correspond to
data taken at quite different altitudes and cutoff rigidities\label{1gev}}

\figure{ Integral Flux at $>$ 2 GeV. Similar to figure 2, but for the
electron detector threshold energy of 2 GeV.\label{2gev}}

\figure{ Integral Flux at $>$ 4 GeV.  Similar to figure 2, but for the
electron detector threshold energy of 4 GeV.\label{4gev}}

\end{document}